\documentclass[11pt]{article}
\usepackage{graphics}
\usepackage{psfrag}
\usepackage{rotating}

\title{Large Direct $CP$ Violation in $B^0 \to \pi^+ \pi^-$ and an Enhanced
Branching Ratio for $B^0 \to \pi^0 \pi^0$}

\author{S. Barshay$^1$, L. M. Sehgal\footnote{{\it E-mail address}: sehgal@physik.rwth-aachen.de (L. M. Sehgal)}$\,\, ^2$  and J. van Leusen$^2$ \\ 
\it{$^1$III. Physikalisches Institut, RWTH Aachen} \\
\it{$^2$Institut f\"ur Theoretische Physik, RWTH Aachen} \\
\it{D-52056 Aachen, Germany}}

\date{}
\begin{document}

\maketitle

\begin{abstract}
Recent measurements of $B^0 \to \pi \pi$ decays reveal two features that are in conflict with
conventional calculations: the channel $B^0 (\overline{B^0}) \to \pi^+ \pi^-$ shows a large
direct $CP$-violating asymmetry, and the channel $B^0 (\overline{B^0}) \to \pi^0 \pi^0$ has an 
unexpectedly high branching ratio. We show that both features can be understood in terms of
strong-interaction mixing of $\pi \pi$ and $D \overline{D}$ channels in the isospin-zero state,
an effect that is important because of the large experimentally observed ratio 
$\Gamma(B^0 / \overline{B^0} \to D^+ D^-) / \Gamma (B^0 / \overline{B^0} \to \pi^+ \pi^-) \approx 50$.
Our dynamical model correlates the branching ratios and the $CP$-violating parameters ${\cal C}$ and ${\cal S}$, for
the decays $B^0 (\overline{B^0}) \to \pi^+ \pi^-$, $B^0 (\overline{B^0}) \to \pi^0 \pi^0$,
$B^0 (\overline{B^0}) \to D^+ D^-$ and $B^0 (\overline{B^0}) \to D^0 \overline{D^0}$.
\end{abstract}

The Belle collaboration has presented new data \cite{BelleRecent} which support their original evidence
\cite{BelleEv} for large direct $CP$ violation in the decays $B^0 (\overline{B^0}) \to \pi^+ \pi^-$, the
asymmetry parameter ${\cal C}$ $( = - {\cal A)}$ being measured to be ${\cal C} = -0.58 \pm 0.15 \pm 0.07$. In a
related development, both the Babar \cite{BaBarLett} and Belle \cite{BelleLee} collaborations have reported
a sizable branching ratio for the decay $B^0 (\overline{B^0}) \to \pi^0 \pi^0$, with an average value
${\rm Br}(B^0 / \overline{B^0} \to \pi^0 \pi^0) = (1.9 \pm 0.6) \times 10^{-6}$. Both of these observations
are unexpectedly large from the standpoint of conventional calculations \cite{BSW,Neubert;Stech,Beneke;Neubert}
based on a short-distance, effective weak Hamiltonian and the assumption of factorization of products of
currents in matrix elements for physical hadron states. In this paper, we carry out a calculation based
upon the idea \cite{Wanninger;Sehgal} of final-state interactions involving the mixing of $\pi \pi$ and $D \overline{D}$
channels. This dynamics provides a natural, correlated explanation of the new experimental facts, and leads to
several further predictions.

To fix notation, we write the three $\overline{B} \to \pi \pi$ amplitudes as
\begin{eqnarray}
A(\overline{B^0} \to \pi^+ \pi^-) & = & N ( \lambda_u a_1 + \lambda_c a_p) \nonumber \\
A(\overline{B^0} \to \pi^0 \pi^0) & = & N ( \lambda_u a_2 - \lambda_c a_p)/ \sqrt{2} \label{FixAmpBpp} \\
A(B^- \to \pi^- \pi^0) & = & N \lambda_u (a_1 + a_2) / \sqrt{2} \nonumber
\end{eqnarray}
Here $a_1$, $a_2$, $a_p$ are, in general, complex numbers and $N$ is a positive normalization factor. The 
parameters $\lambda_u$ and $\lambda_c$ are CKM factors, defined as $\lambda_u = V_{ub} V_{ud}^{\ast}$,
$\lambda_c = V_{cb} V_{cd}^{\ast}$, with magnitudes $| \lambda_u | \cong 3.6 \times 10^{-3}$, 
$| \lambda_c | \cong 8.8 \times 10^{-3}$ and phases given by $\lambda_u = | \lambda_u| e^{- i \gamma}$,
$\lambda_c = - | \lambda_c|$, with $\gamma \approx 60^{\circ}$ \cite{PDG}. The amplitudes in Eq.~(\ref{FixAmpBpp}) are
defined so that their absolute square gives the branching ratio, and they satisfy the isospin relation 
\cite{Gronau;London}
\begin{equation}
\frac{1}{\sqrt{2}} A(\overline{B^0} \to \pi^+ \pi^-) + A(\overline{B^0} \to \pi^0 \pi^0) = A(B^- \to \pi^- \pi^0) \label{IsospinRel}
\end{equation}

From the results of the models discussed in \cite{BSW,Neubert;Stech,Beneke;Neubert}, the parameters appearing in
Eq.~(\ref{FixAmpBpp}) have the following rough representation. The constants $a_1$, $a_2$, $a_p$ are approximately
real (to within a few degrees), with magnitudes $a_1 \approx 1.0$, $a_2 \approx 0.2$, $a_p \approx -0.1$. The 
normalization factor is $N \approx 0.75$; it is here fixed by the empirical branching ratio for $B^- \to \pi^- \pi^0$.
The fact that the parameters $a_1$, $a_2$, $a_p$ are nearly real implies immediately that there is very little direct
$CP$-violating asymmetry between $\overline{B^0} \to \pi^+ \pi^-$ and $B^0 \to \pi^+ \pi^-$, as well as in the channels
$\pi^0 \pi^0$ and $\pi^\pm \pi^0$. Furthermore, the absolute branching ratios following from the above parametrization
are as follows (with experimental values given in parentheses):
\begin{eqnarray}
{\rm Br}(B^\pm \to \pi^\pm \pi^0) = & 5.3 \times 10^{-6} & [{\rm exp.} \, (5.3 \pm 0.8) \times 10^{-6}] \nonumber\\
{\rm Br}(B^0 / \overline{B^0} \to \pi^+ \pi^-) = & 9.2 \times 10^{-6} & [{\rm exp.} \, (4.6 \pm 0.4) \times 10^{-6}] \\
{\rm Br}(B^0 / \overline{B^0} \to \pi^0 \pi^0) = & 0.2 \times 10^{-6} & [{\rm exp.} \, (1.9 \pm 0.6) \times 10^{-6}] \nonumber
\end{eqnarray}
The most striking feature is the strong enhancement of the $\pi^0 \pi^0$ rate compared to this model expectation.

It was pointed out in Ref. \cite{Wanninger;Sehgal} that the $CP$-violating asymmetries and branching ratios in the $B \to \pi \pi$
system would be strongly affected by final-state interactions involving the mixing of the $\pi \pi$ and $D \overline{D}$ channels
in the isospin $I = 0$ state, as a consequence of the large ratio of partial decay widths 
$\Gamma(B^0 \to D^+ D^-)/ \Gamma(B^0 \to \pi^+ \pi^-) \approx \frac{3}{14} |V_{cb}|^2 / |V_{ub}|^2 \approx 26$ expected in the
Bauer-Stech-Wirbel model \cite{BSW}. A large ratio has now been confirmed by the Belle measurement
\cite{BelleBrowder} of the branching ratio ${\rm Br}(B^0 / \overline{B^0} \to D^+ D^-) = 2.5 \times 10^{-4}$, which is
about $50$ times larger than ${\rm Br}( B^0 / \overline{B^0} \to \pi^+ \pi^-)$. This fact gives new urgency to an 
investigation of $\pi \pi \leftrightarrow D \overline{D}$ mixing as a way of resolving the puzzling observations in
$B \to \pi \pi$ decays.

The $\pi \pi$ system exists in the states $I = 0$ or $I = 2$, while the $D \overline{D}$ system has $I = 0$ or $I = 1$. Mixing
can occur between the isospin-zero states
\begin{eqnarray}
| \pi \pi \rangle_0 & = & \sqrt{\frac{2}{3}} | \pi^+ \pi^- \rangle - \sqrt{\frac{1}{3}} | \pi^0 \pi^0 \rangle\\
| D \overline{D} \rangle_0 & = & \sqrt{\frac{1}{2}} \left[ | D^+ D^- \rangle + | D^0 \overline{D^0} \rangle \right] \nonumber
\end{eqnarray}
By contrast, the $I=2$ $\pi \pi$ state and the $I=1$ $D \overline{D}$ state, given by
\begin{eqnarray}
| \pi \pi \rangle_2 & = & \sqrt{\frac{1}{3}} | \pi^+ \pi^- \rangle + \sqrt{\frac{2}{3}} | \pi^0 \pi^0 \rangle\\
| D \overline{D} \rangle_1 & = & \sqrt{\frac{1}{2}} \left[ | D^+ D^- \rangle - | D^0 \overline{D^0} \rangle \right] \nonumber
\end{eqnarray}
are unaffected by mixing. The physical decay amplitudes of $\overline{B^0}$ to the above four states are
\begin{eqnarray}
A_{\pi \pi}^{(0)} & = & \sqrt{\frac{2}{3}} A_{\pi^+ \pi^-} - \sqrt{\frac{1}{3}} A_{\pi^0 \pi^0} \nonumber \\
A_{\pi \pi}^{(2)} & = & \sqrt{\frac{1}{3}} A_{\pi^+ \pi^-} + \sqrt{\frac{2}{3}} A_{\pi^0 \pi^0} \\
A_{D \overline{D}}^{(0)} & = & \sqrt{\frac{1}{2}} \left[A_{D^+ D^-} + A_{D^0 \overline{D^0}} \right] \nonumber \\
A_{D \overline{D}}^{(1)} & = & \sqrt{\frac{1}{2}} \left[A_{D^+ D^-} - A_{D^0 \overline{D^0}} \right] \nonumber
\end{eqnarray}
These physical decay amplitudes are related to the ``bare" amplitudes calculated in the absence of final-state
interactions, i.e. with no mixing, which we denote by $\tilde{A}$:
\begin{eqnarray}
\left( \begin{array}{c} A_{\pi \pi}^{(0)} \\ A_{D \overline{D}}^{(0)} \end{array} \right) & = & S^{\frac{1}{2}} 
\left( \begin{array}{c} \tilde{A}_{\pi \pi}^{(0)} \\ \tilde{A}_{D \overline{D}}^{(0)} \end{array} \right) \nonumber \\
A_{\pi \pi}^{(2)} & = & \tilde{A}_{\pi \pi}^{(2)} \label{AConATilde} \\
A_{D \overline{D}}^{(1)} & = & \tilde{A}_{D \overline{D}}^{(1)} \nonumber
\end{eqnarray}
Here $S$ denotes the strong-interaction $S$ matrix connecting the isospin-zero states $| \pi \pi \rangle_0$ and
$| D \overline{D} \rangle_0$ which can be written generally as\footnote{The two-channel $S$-matrix has been discussed, in
particular in \cite{Donoghue,Suzuki;Wolfenstein}. The $S^{\frac{1}{2}}$ prescription is given in \cite{Neubert;Stech,Donoghue}.
An alternative prescription, using $\frac{1}{2} \left[ {\bf 1} + S \right]$ in place of $S^{\frac{1}{2}}$, has been discussed
by Kamal \cite{Kamal}, and was used in Ref. \cite{Wanninger;Sehgal}.} 
\begin{equation}
S = \left( \begin{array}{cc} \cos 2 \theta \, e^{i 2 \delta_1} & i \sin 2 \theta \, e^{i(\delta_1 + \delta_2)} \\
i \sin 2 \theta \, e^{i(\delta_1 + \delta_2)} & \cos 2 \theta \, e^{i 2 \delta_2}\end{array} \right) \label{Smatrix}
\end{equation}
where $\theta$ is a mixing angle, and $\delta_1$ and $\delta_2$ are the strong-interaction phase shifts for the elastic
scattering of $\pi \pi$ and $D \overline{D}$ systems in the $I=0$ state, at $\sqrt{s} = M_B$. For any choice of these three
parameters, the matrix $S^{\frac{1}{2}}$ can be calculated numerically, and the set of four equations (\ref{AConATilde}) solved
to obtain the physical amplitudes $A_{\pi^+ \pi^-}$, $A_{\pi^0 \pi^0}$, $A_{D^+ D^-}$ and $A_{D^0 \overline{D^0}}$ in terms
of the bare amplitudes. The bare amplitudes are identified with those calculated in the factorization model 
\cite{BSW,Neubert;Stech,Beneke;Neubert}, which we list below
\begin{eqnarray}
\tilde{A}_{\pi^+ \pi^-} & = & N ( \lambda_u a_1 + \lambda_c a_p) \nonumber \\
\tilde{A}_{\pi^0 \pi^0} & = & N ( \lambda_u a_2 - \lambda_c a_p)/\sqrt{2} \label{BareAmp} \\
\tilde{A}_{D^+ D^-} & = & N^{\prime} \lambda_c a_1 \nonumber \\
\tilde{A}_{D^0 \overline{D^0}} & = & 0 \nonumber
\end{eqnarray}
where the first two equations are as in Eq.~(\ref{FixAmpBpp}), and the factor $N^{\prime}$ is determined from the
empirical \cite{BelleBrowder} branching ratio ${\rm Br}(B^0 / \overline{B^0} \to D^+ D^-) = N^{\prime 2} | \lambda_c|^2 a_1^2 = 2.5 \times 10^{-4}$
to be $N^{\prime} = 1.79$.

In order to show, in a transparent way, how the mixing mechanism gives rise to large direct $CP$ violation in $B^0 \to \pi^+ \pi^-$,
as well as an enhanced branching ratio for $B^0 \to \pi^0 \pi^0$, we consider, for illustration, the case where the elastic
phases $\delta_1$ and $\delta_2$ in the $S$ matrix (Eq.~(\ref{Smatrix})) are neglected, so that $S^{\frac{1}{2}}$ may be written as
\begin{equation}
S^{\frac{1}{2}} = \left( \begin{array}{cc} \cos \theta & i \sin \theta \\ i \sin \theta & \cos \theta \end{array} \right) \label{SimpleSroot}
\end{equation}
The amplitudes $A_{\pi^+ \pi^-}$ and $A_{\pi^0 \pi^0}$ for $\overline{B^0}$ decay are then given by
\begin{eqnarray}
A_{\pi^+ \pi^-} & = & \frac{1+ 2 \cos \theta}{3} \tilde{A}_{\pi^+ \pi^-} + \sqrt{2} \frac{1 - \cos \theta}{3} \tilde{A}_{\pi^0 \pi^0} \nonumber \\
& & + i \sin \theta \frac{1}{\sqrt{3}} \left( \tilde{A}_{D^+ D^-} + \tilde{A}_{D^0 \overline{D^0}} \right) \\
A_{\pi^0 \pi^0} & = & \sqrt{2} \frac{1 - \cos \theta}{3} \tilde{A}_{\pi^+ \pi^-} + \frac{2 + \cos \theta}{3} \tilde{A}_{\pi^0 \pi^0} \nonumber \\
& & - i \sin \theta \frac{1}{\sqrt{6}} \left( \tilde{A}_{D^+ D^-} + \tilde{A}_{D^0 \overline{D^0}} \right) \nonumber
\end{eqnarray}
Clearly for $\theta = 0$, the physical amplitudes reduce to the bare amplitudes. Inserting the bare amplitudes from Eq.~(\ref{BareAmp}),
we can rewrite $A_{\pi^+ \pi^-}$ and $A_{\pi^0 \pi^0}$ as linear combinations of $\lambda_u$ and $\lambda_c$:
\begin{eqnarray}
A_{\pi^+ \pi^-} & = & N \left[ \lambda_u \left\{ \frac{1+2 \cos \theta}{3} a_1 + \frac{1- \cos \theta}{3} a_2 \right\} 
+ \lambda_c \left(a_p \cos \theta + a_m \right) \right] \label{AmpofCKM} \\
A_{\pi^0 \pi^0} & = & \frac{N}{\sqrt{2}} \left[ \lambda_u \left\{ \frac{2(1 - \cos \theta)}{3} a_1 + \frac{2+ \cos \theta}{3} a_2 \right\} 
- \lambda_c \left(a_p \cos \theta + a_m \right) \right] \nonumber
\end{eqnarray}
where
\begin{equation}
a_m = i \frac{1}{\sqrt{3}} \sin \theta \frac{N^{\prime}}{N} a_1 \label{Eqforam}
\end{equation}
Note that the isospin relation in Eq.~(\ref{IsospinRel}) continues to be fulfilled.
The important new feature of the amplitudes in Eq.~(\ref{AmpofCKM}) is the appearance of the {\it imaginary} term $a_m$ in the
coefficent of $\lambda_c$, in striking contrast to the real term $a_p$. The imaginary nature of this dynamical term is an inescapable
consequence of $S$-matrix unitarity, which enforces the factor $i$ in the off-diagonal matrix element in Eq.~(\ref{SimpleSroot}).
The term $a_m$, given in Eq.~(\ref{Eqforam}), has a magnitude $|a_m| \approx 1.39 \sin \theta$, and dominates the term $a_p \cos \theta$
even for a modest mixing angle $\sim 0.1$. We will now show that the mixing term $a_m$ has profound consequences for direct $CP$ violation
in the decays $B^0 \to \pi^+ \pi^-$, and for the branching ratio of the channel $B^0 \to \pi^0 \pi^0$.

\section{${\cal C}$ and ${\cal S}$ Parameters for $B^0 \to \pi^+ \pi^-$ and $B^0 \to \pi^0 \pi^0$}

The ${\cal C}$ and ${\cal S}$ parameters derived from the time-dependent asymmetry between $\overline{B^0}$ and $B^0$ decays into
$\pi^+ \pi^-$ are defined as
\begin{eqnarray}
{\cal C}_{+-} & = & \frac{1-| \lambda_{+-} |^2}{1+| \lambda_{+-} |^2} \\
{\cal S}_{+-} & = & \frac{ 2 \, {\rm Im} \lambda_{+-} }{1+| \lambda_{+-} |^2} \nonumber
\end{eqnarray}
where
\begin{equation}
\lambda_{+-} = \frac{q}{p} \frac{A(\overline{B^0} \to \pi^+ \pi^-)}{A(B^0 \to \pi^+ \pi^-)}
\end{equation}
with 
\begin{equation}
\frac{q}{p} = e^{-i 2 \beta}, \, 2 \beta \approx 45^{\circ}. 
\end{equation}

\begin{figure}
\center
\psfrag{theta}[bc][][2]{$\theta$}
\psfrag{Cpm}[tl][][2]{${\cal C}_{+-}$}
\psfrag{Spm}[tl][][2]{${\cal S}_{+-}$}
\makebox[9cm]{\resizebox{9cm}{!}{\includegraphics{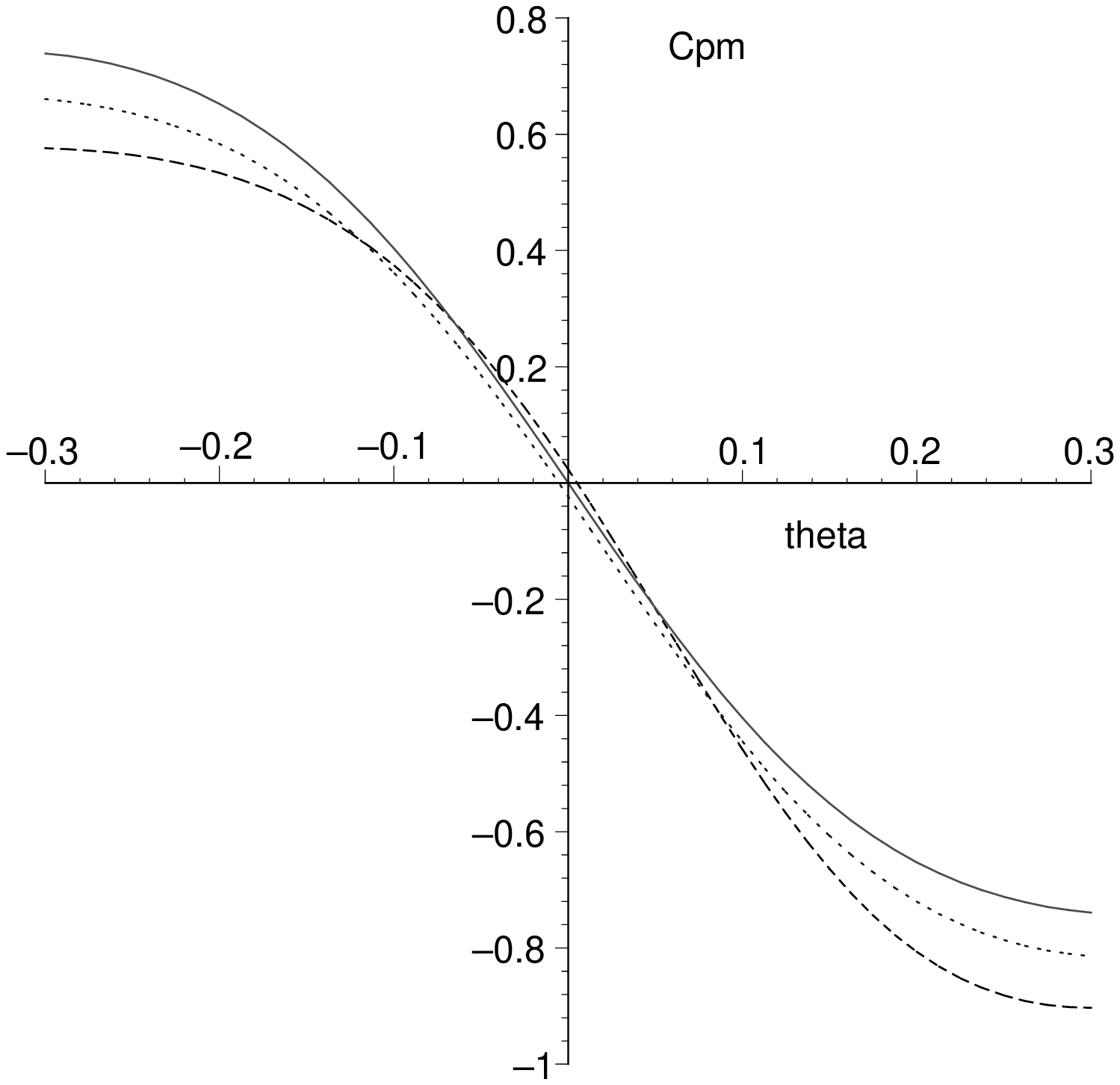}}} \\
\vspace{1cm}
\makebox[9cm]{\resizebox{9cm}{!}{\includegraphics{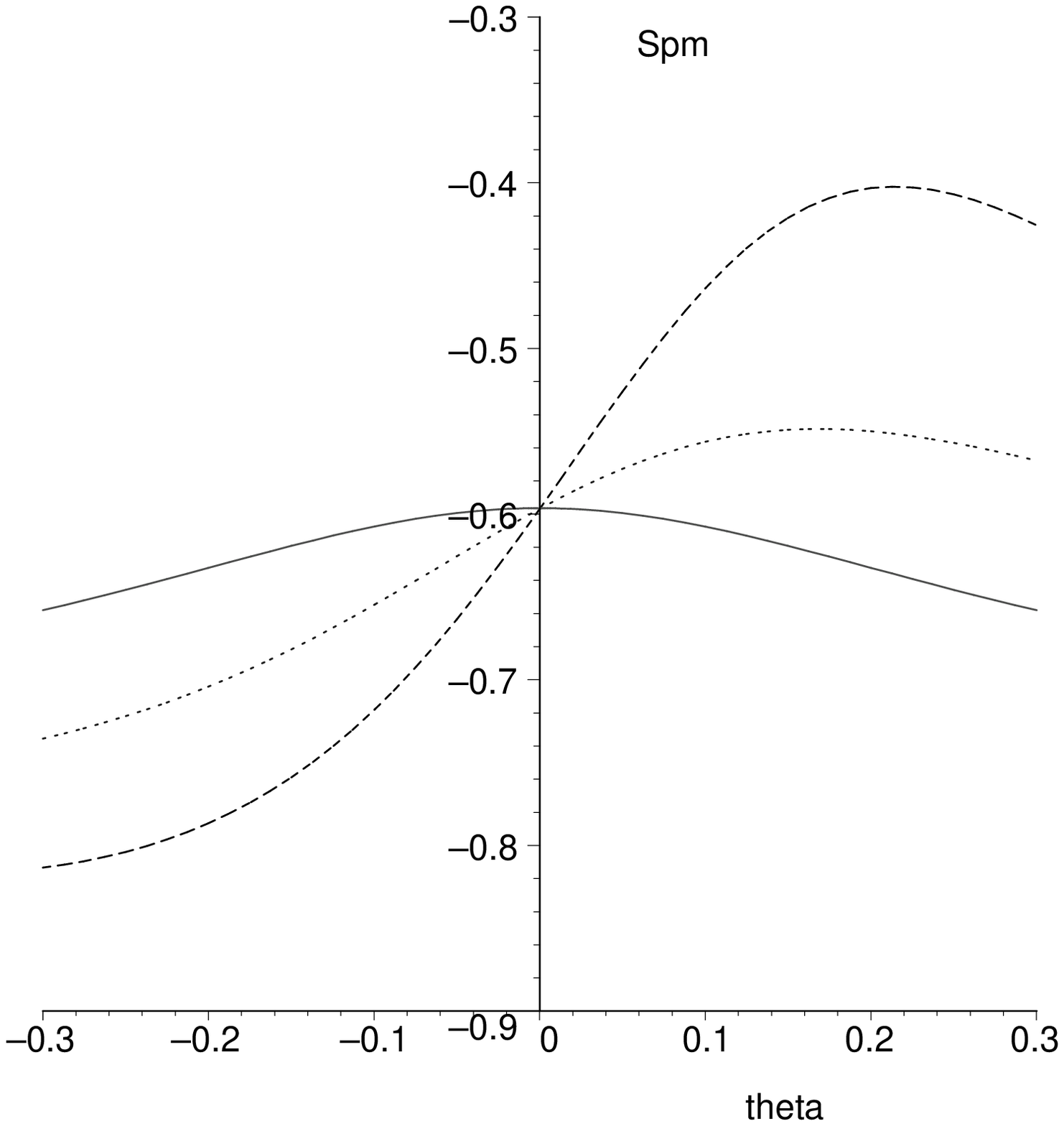}}}
\caption{${\cal C}$ and ${\cal S}$ parameters for the decay $B^0 (\overline{B^0}) \to \pi^+ \pi^-$. Full line is for 
$\delta_1 = \delta_2 = 0^{\circ}$, dotted line for $\delta_1 = -10^{\circ}$, $\delta_2 = -20^{\circ}$,
dashed line for $\delta_1 = 10^{\circ}$, $\delta_2 = -40^{\circ}$.  \label{CpmSpm}}
\end{figure}

\begin{figure}
\center
\psfrag{theta}[bc][][2]{$\theta$}
\psfrag{untgB00}[tl][][1.5]{${\rm Br}(B^0 / \overline{B^0} \to \pi^0 \pi^0)$}
\psfrag{untgBpm}[tl][][1.5]{${\rm Br}(B^0 / \overline{B^0} \to \pi^+ \pi^-)$}
\psfrag{1e\26106}[cc][][1.4]{$1 \times 10^{-6}$}
\psfrag{2e\26106}[cc][][1.4]{$2 \times 10^{-6}$}
\psfrag{3e\26106}[cc][][1.4]{$3 \times 10^{-6}$}
\psfrag{4e\26106}[cc][][1.4]{$4 \times 10^{-6}$}
\psfrag{5e\26106}[cc][][1.4]{$5 \times 10^{-6}$}
\psfrag{6e\26106}[cc][][1.4]{$6 \times 10^{-6}$}
\psfrag{1.0e\26105}[cc][][1.4]{$10 \times 10^{-6}$}
\psfrag{1.2e\26105}[cc][][1.4]{$12 \times 10^{-6}$}
\psfrag{1.4e\26105}[cc][][1.4]{$14 \times 10^{-6}$}
\psfrag{1.6e\26105}[cc][][1.4]{$16 \times 10^{-6}$}
\psfrag{1.8e\26105}[cc][][1.4]{$18 \times 10^{-6}$}
\psfrag{2.0e\26105}[cc][][1.4]{$20 \times 10^{-6}$}
\psfrag{2.2e\26105}[cc][][1.4]{$22 \times 10^{-6}$}
\makebox[9cm]{\resizebox{9cm}{!}{\includegraphics{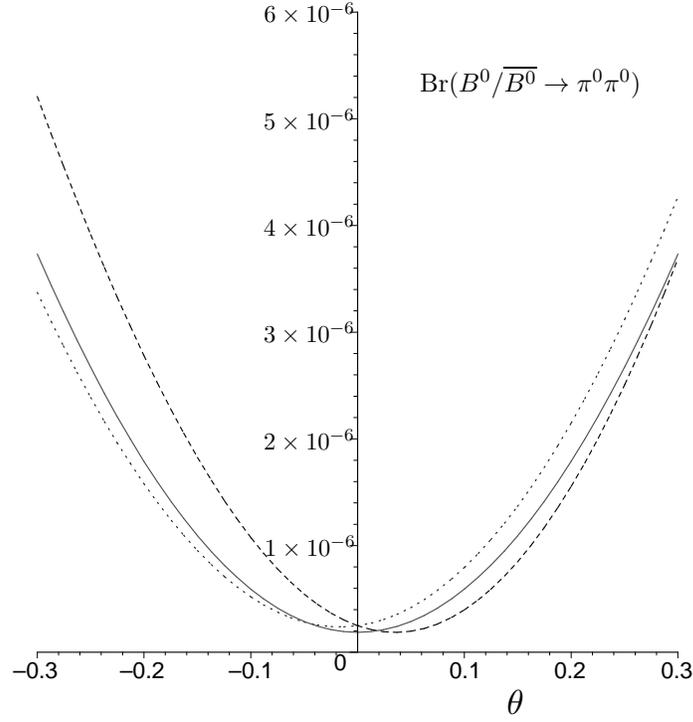}}} \\
\vspace{0.5cm}
\makebox[9cm]{\resizebox{9cm}{!}{\includegraphics{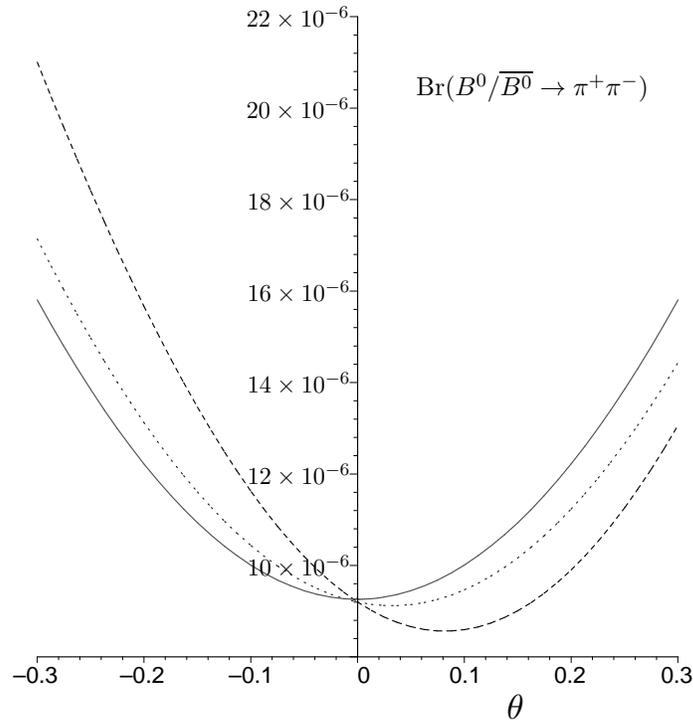}}}
\caption{Average branching ratios for $B^0 / \overline{B^0} \to \pi^0 \pi^0$ and $B^0 / \overline{B^0} \to \pi^+ \pi^-$. 
Full line is for  $\delta_1 = \delta_2 = 0^{\circ}$, dotted line for $\delta_1 = -10^{\circ}$, $\delta_2 = -20^{\circ}$,
dashed line for $\delta_1 = 10^{\circ}$, $\delta_2 = -40^{\circ}$. \label{Bratiospm00}}
\end{figure}

${\cal C}_{+-}$ is the parameter for direct $CP$ violation (i.e. $| A(\overline{B^0} \to \pi^+ \pi^-) / A(B^0 \to \pi^+ \pi^-) | \neq 1$).
Using the amplitude $A_{\pi^+ \pi^-}$ in Eq.~(\ref{AmpofCKM}), we obtain
\begin{equation}
\lambda_{+-} = e^{-i 2 \beta} \left[ \frac{\lambda_u \left( \frac{1+2 \cos \theta}{3} a_1 + \frac{1- \cos \theta}{3} a_2 \right) 
+ \lambda_c \left(a_p \cos \theta + a_m \right)}{\lambda_u^{\ast} \left( \frac{1+2 \cos \theta}{3} a_1 + \frac{1- \cos \theta}{3} a_2 \right) 
+ \lambda_c^{\ast} \left(a_p \cos \theta + a_m \right)} \right]
\end{equation}
The asymmetry parameters ${\cal C}_{+-}$ and ${\cal S}_{+-}$ calculated from the above expression are plotted as functions of $\theta$ in
Fig. \ref{CpmSpm}. Good approximate agreement with data is obtained for $\theta \approx 0.2$ (see Table \ref{TableResults}, where we also list
${\cal C}_{00}$ and ${\cal S}_{00}$). We note that the amplitudes in Eq.~(\ref{AmpofCKM}) have been derived from the matrix
$S^{\frac{1}{2}}$ in Eq.~(\ref{SimpleSroot}), which was obtained from (\ref{Smatrix}) by neglecting the phase shifts $\delta_1$ and $\delta_2$.
We have also explored numerically $S$ matrices with non-zero phases, and indicate in Figs. \ref{CpmSpm} and \ref{Bratiospm00} two examples,
obtained with the values $\delta_1 = \pm 10 ^{\circ}$, $\delta_1+\delta_2 = - 30^{\circ}$. Table \ref{TableResults} gives numerical values
for a few choices of parameters. In all cases, there is a large direct $CP$ violation.

Discussions of the direct $CP$-violating parameter ${\cal C}_{+-}$ are often based on an amplitude for $\overline{B^0} \to \pi^+ \pi^-$
written in the form
\begin{equation}
A_{\pi^+ \pi^-} \sim \left[ e^{-i \gamma} + \frac{P_{\pi \pi}}{T_{\pi \pi}} \right]
\end{equation}
The parametrization in Eq.~(\ref{FixAmpBpp}), based on the models \cite{BSW,Neubert;Stech,Beneke;Neubert}, gives $|P_{\pi \pi} / T_{\pi \pi}| = 0.24$, and
${\rm arg}(P_{\pi \pi}/ T_{\pi \pi}) = 0$. The small phase of the ``penguin-to-tree" ratio $P_{\pi \pi}/T_{\pi \pi}$ is a generic
feature of these models, and is responsible for the prediction ${\cal C}_{+-} \approx 0$,
which is now contradicted by data \cite{BelleRecent}. In our approach, the role of $P_{\pi \pi}/T_{\pi \pi}$ is played by the ratio
\begin{equation}
`` P/T " = - \frac{|\lambda_c| ( a_p \cos \theta + a_m)}{|\lambda_u| \left( \frac{1 + 2 \cos \theta}{3} a_1 + \frac{1- \cos \theta}{3} a_2\right)}
\label{PoverT}
\end{equation}
For a typical value $\theta = 0.2$, this ratio has the modulus $|`` P/T " | \approx 0.77$, and a phase ${\rm arg}(`` P/T ") \approx - 70^{\circ}$.
The difference is a consequence of the term $a_m$ in Eq.~(\ref{PoverT}), which reflects the physical final-state interaction of the $\pi \pi$ system,
as implemented in our model through $\pi \pi \leftrightarrow D \overline{D}$ mixing.

\section{Branching Ratio for $B^0 \to \pi^0 \pi^0$ and $B^0 \to \pi^+ \pi^-$}

The branching ratios (averaged over $B^0$ and $\overline{B^0}$) may be calculated in our model by taking the absolute square of the
$\overline{B^0}$ decay amplitudes in Eq.~(\ref{AmpofCKM}), and the corresponding amplitudes for $B^0$ decay. The results are shown in Fig.
\ref{Bratiospm00}. It is remarkable that the empirical branching ratio for $B^0 / \overline{B^0} \to \pi^0 \pi^0$ is accurately reproduced, using
the same value $\theta \approx 0.2$ which accounts for the asymmetry parameter ${\cal C}_{+-}$. We also note that the branching ratio
$B^0 / \overline{B^0} \to \pi^+ \pi^-$ remains close to its bare value, and can be lowered slightly with the introduction of phases
$\delta_1$ and $\delta_2$. Numerical results for ${\rm Br}(B^0 / \overline{B^0} \to \pi^0 \pi^0)$ and ${\rm Br}(B^0 / \overline{B^0} \to \pi^+ \pi^-)$
are listed in Table \ref{TableResults}.

\section{Branching Ratio for $B^0 \to D^0 \overline{D^0}$}

Since our model treats the $\pi \pi$ and $D \overline{D}$ states with $I=0$ as a coupled system, it also produces predictions for branching
ratios and asymmetry parameters in $B^0 \to D^+ D^-$ and $B^0 \to D^0 \overline{D^0}$. The amplitudes after mixing are
\begin{eqnarray}
A_{D^+ D^-} & = & \frac{1}{2} \left[ i \sin \theta \sqrt{\frac{2}{3}} \left( \sqrt{2} \tilde{A}_{\pi^+ \pi^-} - \tilde{A}_{\pi^0 \pi^0} \right) \right. \nonumber \\
& & \left. + ( \cos \theta + 1) \tilde{A}_{D^+ D^-} + (\cos \theta -1) \tilde{A}_{D^0 \overline{D^0}} \right] \label{AmpofDD} \\
A_{D^0 \overline{D^0}} & = & \frac{1}{2} \left[ i \sin \theta \sqrt{\frac{2}{3}} \left( \sqrt{2} \tilde{A}_{\pi^+ \pi^-} - \tilde{A}_{\pi^0 \pi^0} \right) \right. \nonumber \\
& & \left. + ( \cos \theta - 1) \tilde{A}_{D^+ D^-} + (\cos \theta +1) \tilde{A}_{D^0 \overline{D^0}} \right] \nonumber
\end{eqnarray}
Of particular interest is the branching ratio for $B^0 / \overline{B^0} \to D^0 \overline{D^0}$, since it vanishes at the level of the bare
amplitude ($\tilde{A}_{D^0 \overline{D^0}} = 0$), and is induced by mixing with the $\pi \pi$ system. For $\theta = 0.2$, ignoring the phases
$\delta_1$, $\delta_2$, our model predicts
\begin{equation}
{\rm Br}( B^0 / \overline{B^0} \to D^0 \overline{D^0}) = 1.45 \times 10^{-7}
\end{equation}
(At this low level, one must assume that other sources of final-state interaction or a non-zero bare amplitude could raise this branching ratio
further.)
Direct $CP$ violation follows from $A_{D^0 \overline{D^0}}$ in Eq.~(\ref{AmpofDD}): ${\cal C}_{D^0 \overline{D^0}} = -0.50$ for $\theta=0.2$.
Direct $CP$ violation in $D^+ D^-$ (and in $\pi^- \pi^0$) is small, because  these decays are dominated by a single amplitude. There is
little mixing in $A_{D^+ D^-}$ in Eq.~(\ref{AmpofDD}) (and none in the $I=2$ amplitude for $\pi^- \pi^0$).

To conclude, we have demonstrated a mechanism of final-state interactions among physical hadrons in $B^0 \to \pi \pi$ decays which predicts
a large direct $CP$-violating parameter ${\cal C}_{+-}$. The same mechanism enhances the theoretical prediction for the branching ratio of
$B^0 / \overline{B^0} \to \pi^0 \pi^0$ to the experimentally observed level. Predictions are made for the ${\cal C}$ and ${\cal S}$
parameters of $B^0 (\overline{B^0}) \to \pi^0 \pi^0$ decays, and for the branching ratio of $B^0 / \overline{B^0} \to D^0 \overline{D^0}$.
The model makes essential use of the
large empirical ratio $\Gamma (B^0 / \overline{B^0} \to D^+ D^-) / \Gamma (B^0 / \overline{B^0} \to \pi^+ \pi^-) \approx 50$.
Its success in the present context leads to the expectation that sizable direct $CP$ violation could be observed in other charmless $B$
decays, in which an amplitude of order $\lambda_u$ receives a dynamical contribution proportional to $\lambda_c$, through
mixing with a channel possessing a large branching ratio. The resulting amplitude contains two pieces which are comparable in magnitude
and have different weak-interaction and strong-interaction phases. We have treated earlier \cite{Barshay;Rein;Sehgal} the charged-particle
decays $B^\pm \to \eta \pi^\pm$ (and $B^\pm \to \eta^{\prime} \pi^\pm$), which are influenced by mixing with the channel 
$B^\pm \to \eta_c \pi^\pm$, and have predicted significant direct $CP$ violation. Evidence for a sizable violation in
$B^\pm \to \eta \pi^\pm$ has indeed been reported in one experiment \cite{BabarHep}, the first ever seen in a charged-particle decay.

\begin{sidewaystable}
\begin{tabular}{|c|c|c|c|c|cl|}
\hline
& No & \multicolumn{3}{c|}{With mixing} & & \\ 
\cline{3-5}
Observable & mixing & $\theta = 0.2$ & $\theta = 0.17$ & $\theta = 0.2$ & \multicolumn{2}{c|}{Data} \\
& & $\delta_1 = 0^{\circ}$ & $\delta_1 = -10^{\circ}$ & $\delta_1 = 10^{\circ}$ & & \\
& & $\delta_2 = 0^{\circ}$ & $\delta_2 = -20^{\circ}$ & $\delta_2 = -40^{\circ}$ & & \\
\hline
${\cal C}_{+-}$ & $\pm 0.00$ & $-0.65$ & $-0.66$ & $-0.81$ & $-0.58 \pm 0.15 \pm 0.07$ & (Belle \cite{BelleRecent}) \\
& & & & & $-0.30 \pm 0.25 \pm 0.04$ & (Babar \cite{BabarTab}) \\
${\cal S}_{+-}$ & $-0.60$ & $-0.63$ & $-0.55$ & $-0.40$ & $-1.00 \pm 0.21 \pm 0.07$ & (Belle \cite{BelleRecent}) \\
& & & & & $+0.02 \pm 0.34 \pm 0.05$ & (Babar \cite{BabarTab}) \\
${\rm Br}(B^0 / \overline{B^0} \to \pi^0 \pi^0)$
 & $0.2$ & $1.8$ & $1.7$ & $1.6$ & $1.7 \pm 0.6 \pm 0.2$ & (Belle \cite{BelleLee}) \\
& & & & & $2.1 \pm 0.6 \pm 0.3$ & (Babar \cite{BaBarLett}) \\
${\rm Br}(B^0 / \overline{B^0} \to \pi^+ \pi^-)$ 
& $9.3$ & $12.2$ & $10.5$ & $9.9$ & $4.4 \pm 0.6 \pm 0.3$ & (Belle \cite{Tomura}) \\
& & & & & $4.7 \pm 0.6 \pm 0.2$ & (Babar \cite{BabarTab}) \\
& & & & & $4.5_{-1.2 -0.4}^{+1.4 + 0.5}$ & (CLEO \cite{CLEO}) \\
\hline
${\cal C}_{00}$ & $\pm 0.00$ & $+0.48$ & $+0.51$ & $+0.56$ & & \\
& & & & & &\\
${\cal S}_{00}$ & $+0.73$ & $-0.65$ & $-0.78$ & $-0.49$ & & \\
\hline
\end{tabular}
\caption{Observables for different mixing angles $\theta$ and strong-interaction phases $\delta_1$ and $\delta_2$.
All branching ratios are given in units of $10^{-6}$. \label{TableResults}}
\end{sidewaystable}

\end{document}